\documentclass[manuscript]{aastex}

\usepackage{amsmath}
\usepackage{amssymb}
\usepackage{graphicx}
\usepackage{epstopdf}

\shorttitle{PGCC G108.84-00.81}
\shortauthors{Kim et al.}

\begin{document}

\newcommand{\tco}{CO 1--0}
\newcommand{\thco}{$^{13}$CO 1--0}
\newcommand{\ceo}{C$^{18}$O 1--0}
\newcommand{\ceom}{C$^{18}$O}
\newcommand{\tcot}{CO 2--1}
\newcommand{\thcot}{$^{13}$CO 2--1}
\newcommand{\ceot}{C$^{18}$O 2--1}
\newcommand{\hcop}{HCO$^+$ 1--0}
\newcommand{\nhp}{N$_2$H$^+$ 1--0}
\newcommand{\nhpm}{N$_2$H$^+$}
\newcommand{\hcn}{HCN 1--0}
\newcommand{\nht}{NH$_{3}$ (1,1)}
\newcommand{\nhtm}{NH$_{3}$}
\newcommand{\hcopf}{HCO$^+$ 4--3}
\newcommand{\sou}{PGCC G108.84--00.81}
\newcommand{\m}{\micron}
\newcommand{\cd}{$N(\textrm{H}_2$)}
\newcommand{\cdc}{$N(\textrm{C$^{18}$O}$)}
\newcommand{\atomh}{H$_2$}
\newcommand{\nd}{$n({\textrm{H}}_{2})$}
\newcommand{\tex}{$T_{\textrm{ex}}$}
\newcommand{\td}{$T_{\textrm{dust}}$}
\newcommand{\tkin}{$T_{\textrm{kin}}$}
\newcommand{\snr}{SNR G109.1--1.0}

\title{Star formation conditions in a Planck Galactic Cold Clump, G108.84-00.81}

\author{Jungha Kim\altaffilmark{1, 2, 3}, Jeong-Eun Lee\altaffilmark{1}, Tie Liu\altaffilmark{4}, Kee-Tae Kim\altaffilmark{4}, Yuefang Wu\altaffilmark{5}, Ken$'$ichi, Tatematsu\altaffilmark{2, 3}, Sheng-Yuan Liu\altaffilmark{6}, JCMT Large Program ``SCOPE" collaboration, and TRAO Key Science Program ``TOP" collaboration}
\affil{$^1$ School of Space Research, Kyung Hee University, 1732, Deogyeong-daero, Giheung-gu, Yongin-si, Gyeonggi-do, Korea \\
          $^2$ Department of Astronomical Science, SOKENDAI (The Graduate University for Advanced Studies), 2-21-1 Osawa, Mitaka, Tokyo 181-8588, Japan\\    
          $^3$ National Astronomical Observatory of Japan, 2-21-1 Osawa, Mitaka, Tokyo 181-8588, Japan \\
	   $^4$ Korea Astronomy and Space Science Institute, 776 Daedeokdaero, Yuseong, Daejeon 34055, Korea\\
	   $^5$ Department of Astronomy, Peking University, 100871, Beijing, China \\
	   $^6$ Academia Sinica, Institute of Astronomy and Astrophysics, P.O. Box 23-141, Taipei 106, Taiwan  }
	
\begin{abstract}
We present the results from a series of ground-based radio observations toward a Planck Galactic Cold Clump (PGCC), \sou, which is located in one curved filamentary cloud in the vicinity of an extended H{\sc ii} region Sh2-152 and SNR G109.1-1.0.
\sou\ is mainly composed of two clumps, ``G108--N'' and ``G108--S''.
In the 850 \m\ dust continuum emission map, G108--N is shown as one component while G108--S is fragmented into four components.
There is no infrared source associated with G108--N while there are two infrared sources (IRS 1 and IRS 2) associated with G108--S.
The total mass of G108--N is larger than the jeans mass, suggesting that G108--N is gravitationally unstable and a potential place for a future star formation.
The clump properties of G108--N and G108--S such as the gas temperature and the column density, are not distinctly different.
However, G108--S is slightly more evolved than G108--N, in the consideration of the CO depletion factor, molecular abundances, and association with infrared sources.
G108--S seems to be affected by the compression from Sh2-152, while G108--N is relatively protected from the external effect. 
\clearpage
\end{abstract}

\section{INTRODUCTION}

Expanding H{\sc ii} regions or supernova explosions can strongly influence their surrounding interstellar medium (ISM) and regulate star formation. The shocks induced
in H{\sc ii} regions or supernova remnants may gather surrounding molecular gas to form dense shells, which may collapse to form stars later \citep{el77,whit94}.
This so-called ``collect and collapse" process can self-propagate and lead to sequential star formation \citep{el77,whit94}. The ``collect and collapse" model was firstly developed to explain the age sequence of spatially
distinct OB subgroups in nearby OB associations such as Orion OB1 \citep{el77}. More and more evidences in recent observations indicate that new generation of stars can form in shells or pillar structures surrounding H{\sc ii} regions (see \citealt{liu12a,dale15,liu15,liu16a} and the references therein). \cite{thom12} estimated that the fraction of massive
stars in the Milky Way formed by triggering process could be between 14 and 30 $\%$ through statistical studies of young stellar objects (YSOs) projected against the rims of Spitzer infrared bubbles.

However, recent numerical simulations suggest that stellar feedback from massive stars always results
in a lower star formation efficiency and most signs such as the ages and geometrical distribution of stars relative to the feedback
source or feedback-driven structure (e.g. shells, pillar structures), may not be substantially helpful in distinguishing triggered star formation from spontaneous star formation \citep{dale12,dale15}.  
Studying the conditions (e.g., temperature, density, velocity, chemistry) of dense cores near H{\sc ii} regions may be helpful to distinguish triggered star formation from spontaneous star formation. 
The dense cores should be externally heated and compressed by H{\sc ii} regions if their star formation would be triggered.

The \textit{Planck} satellite survey provides a catalog containing 13,188 Planck Galactic cold clumps \citep[PGCCs;][]{planck15}.
The clumps have dust temperatures lower than 14 K, indicating that PGCCs correspond to the coldest portion of the ISM.
A survey in the $J$ = 1--0 lines of $^{12}$CO, $^{13}$CO, and C$^{18}$O toward 674 PGCCs, has been carried out using the 13.7 m telescope of Purple Mountain Observatory (PMO) \citep{wu12,liu13,zhang16}.
Nearly 98 $\%$ of the clumps have the excitation temperatures ($T_{\textrm{ex}}$) of \tco\ lower than 16 K.
Clumps with excitation temperatures above 16 K are located in star-forming regions, suggesting that the high \tex\ is related to star-forming activities \citep{wu12}.
A large fraction of PGCCs have lower temperatures than the typical temperature of Infrared dark clouds (IRDCs) \citep[15 K;][]{pillai06}, which are known to be the earliest phase of star formation \citep{carey98,simon06}.
It suggests that PGCCs may represent an earlier evolutionary stage than IRDCs.
To investigate the initial conditions of star formation, and especially to study the environmental effect on dense core formation, a legacy survey of PGCCs using the Submillimeter Common User Bolometer Array-2 (SCUBA-2) on board of the James Clerk Maxwell Telescope (JCMT) at East Asia Observatory, ``SCUBA-2 Continuum Observations of Pre-protostellar Evolution (SCOPE\footnote{https://topscope.asiaa.sinica.edu.tw/tiki/tiki-index.php})" is in progress (liu et al. 2017, in preparation).

\sou\ is observed as part of SCOPE.
It has an excitation temperature of $\sim$15 K and \atomh\ column density of 2.0 $\times 10^{22}$ cm s$^{-1}$ \citep{wu12}.
Hereafter we call the region covered by PMO observations (500\arcsec $\times$ 500\arcsec) as \sou\ and the central part of \sou\ covered by SCUBA-2 as G108.
A supernova Remnant (SNR) G109.1--1.0 and an H{\sc ii} region, Sh2-152, are located near \sou. 
Therefore, \sou\ is an ideal target to investigate the environmental effect on the initial conditions of star formation.

In this study, we investigate the interaction between \sou\ and its surrounding environments with CO line observations.
The chemical and physical properties of G108 are also studied in detail through a series of observations in both continuum and molecular lines.
In Section 2, our observations toward \sou\ and G108 are summarized.
Results and analysis are presented in Section 3 and Section 4, respectively.
We discuss the environmental influence on the star formation in \sou\ in Section 5.
Finally, a summary of the study is given in Section 6.

\section{THE DATA}

A series of observations toward \sou\ have been conducted with ground-based radio telescopes.
The reduction process of molecular line data was done using the GILDAS/CLASS\footnote{see http://www.iram.fr/IRAMFR/GILDAS/doc/html/class-html/class.html.} package.
A summary of the molecular line observations is in Table 1.

\subsection{PMO Observations}
The \tco, \thco, and \ceo\ line mapping observations of \sou\ were carried out using the 13.7 m radio telescope of the PMO at De Ling Ha in 2011 May.
The On-The-Fly (OTF) observing mode was applied.
More details of the OTF observations with the PMO 13.7 m telescope toward Planck cold clumps can be found in \citet{liu12}.
The beam Full Width at Half Maximum (FWHM) is $\sim$52$\arcsec$ at 115 GHz and the beam efficiency is $\sim$0.45.
Newly installed nine-beam array receiver in the single-sideband mode was used.
The Fast Fourier Transform Spectrometer (FFTS), which has 16,384 channels in a total bandwidth of 1 GHz, were used.
The velocity resolution of the \tco\ line is 0.16 km s$^{-1}$ and that of both \thco\ and \ceo\ is 0.17 km s$^{-1}$.
\tco\ was observed at the upper sideband and \thco\ and \ceo\ were observed simultaneously at the lower sideband.
The map size for further analysis is about 480\arcsec$\times$480\arcsec.

\subsection{CSO Observations}

Molecular line observations of \sou\ were carried out with the 10 m telescope of the Caltech Submillimeter Observatory (CSO) in 2014 January (see Meng et al. in preparation).
We mapped the \tcot, \thcot, and \ceot\ lines toward G108.
The Sidecab receiver and the FFTS2 spectrometer were used.
The velocity resolution is about 0.35 km s$^{-1}$.
The beam FWHM is 32.5$\arcsec$ and the beam efficiency is 0.76 at 230 GHz.
A map area of 120\arcsec$\times$120\arcsec\ centered at G108 is selected for further analysis because of apparent higher noise levels at map edges.

\subsection{NRO Observations}
The northern part and the southern part of G108 were observed separately with the 45 m radio telescope of the Nobeyama Radio Observatory (NRO) in 2015 December.
We observed \nhp\ with the TZ1 receiver.
The velocity resolution is about 0.2 km s$^{-1}$.
The beam FWHM at 93 GHz is 20.4$\arcsec$ and the beam efficiency is 0.54.
Each mapped area toward the northern and southern part of G108 is 50\arcsec$\times$50\arcsec.

\subsection{IRAM Observations}
We also observed the \hcop\ and \hcn\ lines toward the northern part of G108 with the 30 m radio telescope of the Institut de Radioastronomie Millim\'{e}trique (IRAM) in 2014 May.
The velocity resolution is 0.33 km s$^{-1}$.
The beam FWHM and beam efficiency are $\sim$29 $\arcsec$ and 0.81, respectively.
The map size is 80 \arcsec$\times$80 \arcsec.

\subsection{Effelsberg Observations}
Observations of the northern part of G108 in the \nht\ line was carried out with the 100 m radio telescope of Effelsberg in 2015 April.
The \nht\ line was observed with 1.3 cm double beam secondary focus receiver.
The beam FWHM and beam efficiency are $\sim$38 $\arcsec$ and 0.79, respectively.
The velocity resolution is 0.15 km s$^{-1}$.
An area of 80 \arcsec$\times$80 \arcsec\ was mapped.

\subsection{KVN Observations}
Single-point observations of the \hcn\ line were carried out toward the peak positions of the 850 \m\ dust continuum map in 2016 June with the Korean VLBI Network (KVN) 21 m telescopes \citep{kim11}.
Observations were conducted at the three stations of KVN (Yeonsei, Ulsan, and Tamna).
We used the digital spectrometer with 32 MHz in bandwidth divided into 4096 channels, providing velocity resolution of $\sim$0.2 km s$^{-1}$ at 88 GHz.
The averaged beam FWHM and beam efficiency at the three stations are $\sim$32$\arcsec$ and $\sim$0.39 at 88 GHz, respectively.

\subsection{JCMT Observations}
The 850 \m\ continuum emission map was obtained with JCMT/SCUBA-2 in 2015 May in the "CV Daisy" mapping mode, which is optimised for point sources.
The CV Daisy is designed for small compact sources providing a deep 3$\arcmin$ region in the centre of the map but coverage out to beyond 12$\arcmin$ \citep{bint14}.  
All the SCUBA-2 850 \m\ continuum data were reduced using an iterative map-making technique \citep{chap13}. 
Specifically the data were all run with the reduction tailored for compact sources, filtering out scales larger than 200$\arcsec$ on a 4$\arcsec$ pixel scale.
The beam FWHM of SCUBA-2 at 850 \m\ is $\sim$13 $\arcsec$.
The field of view of SCUBA-2 is about 8 \arcmin\ at 850 \m.
However, an area of about 5\arcmin$\times$5\arcmin\ entered at \sou\ is used for further analysis.

Molecular line observations in \hcopf\ toward the northern part of G108 were also carried out using the JCMT Heterodyne Array Receiver Program (HARP) system in 2015 September.
HARP is a Single Sideband array receiver composed of 16 SIS mixers.
The beam FWHM at 350 GHz is $\sim$14 $\arcsec$ and the beam efficiency is 0.64.
The mapping area is about 50 \arcsec$\times$50 \arcsec.
We used the "ORAC-DR" pipeline in STARLINK for imaging data reduction.

\subsection{Infrared Data}

We also included images from \textit{Wide-field Infrared Survey Explorer} (WISE) \citep{wright10}.
We gathered available photometric data of Young Stellar Object (YSO) candidates in G108 from several catalogs.
The data at 3.4, 4.6, 12, and 22 \m\ are from the ALLWISE catalog including 2MASS data.
The data at 9 and 18 \m\ are from the AKARI/IRC Point Source Catalogue \citep[AKARIPSC;][]{ishihara10}.

\section{RESULTS}

\subsection{Filament structure in the \tco\ and \thco\ maps}

\sou\ is surrounded by dynamically active regions such as a supernova remnant, SNR G109.1--1.0, and an H{\sc ii} region, Sh2-152.
According to Planck Early Cold Clump catalog note\footnote{http://irsa.ipac.caltech.edu/Missions/planck.html}, there is one IRAS point source, IRAS 22565+5839, associated with \sou\ within 5 arcmin from the center of \sou.
One large filament structure is shown in the integrated intensity map of \tco\ (Figure 1).

This filament structure is largely divided into three parts: the northeast region associated with a star \citep{yung14}, IRAS 22576+5843, the southern region directly linked to Sh2-152, and the central part defined as G108, containing IRAS 22565+5839.
As shown in the \thco\ map in Figure 2, G108 is composed of two components.
One is located at northeast (NE) from the central position and the other is located at southwest (SW).
We use ``G108--N'' to indicate the NE component and ``G108--S'' for the SW component.

The systemic velocity of --49.6 km s$^{-1}$ has been measured by Gaussian fitting the \ceo\ spectra toward G108.
The nearby H{\sc ii} region, Sh2-152, has a similar systemic velocity to G108.
We adopt the distance to \sou\ as 3.21 $\pm$ 0.21 kpc from the distance estimation of Sh2-152 by \citet{ramirez11} based on near-infrared extinction.
According to the moment 1 maps presented in Figure 2, a small velocity gradient is present at the SW part of \sou.

\subsection{Two clumps within G108}

Figure 3 presents the integrated intensity maps of \tcot\ and \thcot\ overlaid onto their Moment 1 maps obtained at the CSO.
G108--N and G108--S seem separated in the \tcot\ map.
The integrated intensity map of \thcot\ presents more fragmented structures than that of \tcot.
The component with a blue-shifted velocity with regard to the systemic velocity ($\sim$ --50.5 km s$^{-1}$), is shown in the west of G108--S.
The red-shifted velocity components are located along the eastern edge of filament in the \thcot\ map.

The spectra of CO isotopologues obtained with the CSO telescope toward G108--N and G108--S are shown in Figure 4.
The central velocities of three lines are coincident with each other.
Figure 5 shows the integrated intensity ratios, \tcot$/$\tco\ and \thcot$/$\thco.
The enhancement of \tcot$/$\tco\ ratios ($> 1$) are known toward molecular clouds associated with SNRs \citep{seta98}.
However, in G108, the \thcot$/$\thco\ ratio is the most enhanced at the southern edge of G108--S, which is close to the H{\sc ii} region.

In Figure 6, the 850 \m\ dust continuum emission map is overlaid onto the WISE 12 \m\ image.
Dust continuum emission toward G108 is also largely divided into two parts (G108--N and G108--S), as molecular line emission.
Distribution of dust continuum emission of G108--N shows a cometary structure.
The head is located at the NE from the center with a tail structure extended to SW.
Highly fragmented structure appears in G108--S.
At least four components exist in G108--S.
G108--S1 to S4 are assigned along with Right Ascension.
As shown in the integrated intensity maps of \tcot, \thcot, and \ceot\ overlaid onto the 850 \m\ dust continuum emission map (Figure 7), the intensity peaks of all three lines are off from that of the dust continuum.

\subsection{A starless clump, G108--N}
To investigate the molecular environments of the starless clump, G108--N, single-dish observations were conducted.
The integrated intensity maps of \hcop\ and \hcn\ obtained with the IRAM 30 m telescope are shown in Figure 8.
The \hcop\ and \hcn\ show a similar distribution to each other and also to the distribution of 850 \m\ continuum emission.
The peak positions of \hcop\ and \hcn\ are slightly shifted from that of the 850 \m\ continuum emission peak.
The emission of the two lines distributes along the cometary structure in the NE-SW direction.
The head at the NE with a tail extended to SW is shown in \hcop\ and \hcn\ maps while \hcopf\ obtained with the JCMT/HARP shows a compact distribution near the head part (Figure 8c).
This cometary structure is also shown in the integrated intensity map of \nht\ obtained with the Effelsberg 100 m telescope (Figure 9).

The line profiles of all observed molecular lines toward the 850 \m\ dust continuum peak of G108--N are presented in Figure 10.
The centroid velocities of all lines are consistent.

\section{ANALYSIS}

\subsection{IR source in association with G108}
G108--N has no associated infrared (IR) source.
On the other hand, two IR sources associated with G108--S1 and G108--S2 are detected at all four bands of WISE (Figure 6).
Hereafter, the IR sources associated with G108--S1 and G108--S2 are called IRS 1 and IRS 2, respectively.
IRAS 22565+5839, an associated IRAS point source from the PGCC catalog, is located at the center of the projected separation of IRS 1 and IRS 2, suggesting that the IRAS source is resolved at the WISE 12 \m\ with a higher spatial resolution ($\sim$6 \arcsec).
IRS 1 and IRS 2 are identified as YSO candidates based on the diagnostic color-color and color-magnitude diagrams (See Figure 11) in the YSO identification schemes presented by \citet{koenig14}.

The evolutionary stages of IRS 1 and IRS 2 are classified by their spectral indices and bolometric temperatures.
A spectral index $\alpha$ is derived by fitting the photometric data in the wavelength range from 2 to 24 \m\ as the following \citep{evans09};
\begin{equation}
\alpha = \frac{d\ \textrm{log}(\lambda S(\lambda))}{d\ \textrm{log}(\lambda)}.
\end{equation}
Here, $S(\lambda)$ is the flux density at the wavelength $\lambda$.
The photometric data used to estimate the spectral index are listed in Table 2.
The spectral indices of IRS 1 and IRS 2 are 0.61 and 0.48, respectively.
Using the classification scheme from \citet{evans09} (i.e., Class I for 0.3 $\leq \alpha$, Flat for --0.3 $\leq \alpha <$ 0.3, Class II for --1.6 $\leq \alpha <$ --0.3, and Class III for $\alpha <$ --1.6), both are classified as Class I sources.

The bolometric luminosities ($L_{\textrm{bol}}$) and bolometric temperatures ($T_{\textrm{bol}}$) of the two YSO candidates are also calculated.
The $L_{\textrm{bol}}$ of IRS 1 and IRS 2 are 27 and 30 $L_{\sun}$, respectively, indicating these two sources are low-mass YSOs ($L_{\textrm{bol}} <$ 50 L$_{\sun}$).
IRS 1 and IRS 2 have $T_{\textrm{bol}}$ of 341 K and 614 K, respectively.
Following the classification criteria from \citet{myers93} (i.e., Class 0 for $T_{\textrm{bol}} < 70$ K, Class I for 70 K $\leq T_{\textrm{bol}} \leq$ 650 K, and Class II for 650 K $< T_{\textrm{bol}} \leq$ 2800 K), both are also classified as Class I sources.
The CO outflows, which are commonly associated with embedded Class 0/I sources, are not detected in these two Class I sources.
Higher resolutions and better sensitivities may reveal the associated outflows.
The bolometric luminosities, the bolometric temperatures, and the spectral indices of the two IR sources are listed in Table 3.

\subsection{Clump properties of G108--N and G108--S}

\subsubsection{Gas temperature}

We assume that the $^{12}$CO line is optically thick and all levels are in local thermodynamic equilibrium (LTE) to derive its excitation temperature \citep{liu13}.
Therefore, the excitation temperatures ($T_{\textrm{ex}}$) in all levels and the kinetic temperature ($T_{\textrm{kin}}$) are the same.
We also assume that dust and gas are well coupled ($T_{\textrm{ex}} = T_{\textrm{kin}} = T_{\textrm{dust}}$).

The excitation temperature is derived from the brightness temperature of \tcot, which is optically thick ($\tau_{\nu} \gg 1$).
The expression for the brightness temperature, $T_{\textrm{b}}$ is:
\begin{equation}
T_{\textrm{b}}=[J_\nu(T_{\textrm{ex}})-J_{\nu}(T_{\textrm{bg}})](1-e^{-\tau_{\nu}})f,
\end{equation}
where $J_{\nu}=h\nu_{\textrm{u}}/k(e^{h\nu_{\textrm{u}}/kT} -1)^{-1}$, $\tau_{\nu}$ is optical depth, and $f$ is the beam filling factor and assumed as unity.

We compared the calculated \tex\ with the result of the previous studies.
The mean $T_{\textrm{ex}}$ toward G108--N and G108--S are 13.4 K and 12.5 K, respectively, which are consistent with the \td\ (11.9 $\pm$ 2.8 K) of \sou\ from the Planck PGCC catalog \citep{planck15}.
\tkin\ from the ammonia observations \citep{tatematsu17} of G108--N and G108--S are 14.3 K and $<$ 19.1 K, respectively.
The calculated \tex\ is also consistent with this \tkin, suggesting that our LTE assumption is reliable.

\subsubsection{\atomh\ Column Density derived from $C^{\textit{18}}O$ 2-1 line}
We calculate the \atomh\ column density from the C$^{18}$O line assuming LTE and the same excitation temperature as CO.
The column density, N$_{\textrm{thin}}$, under the assumption of optically thin emission ($\tau \ll$ 1) can be estimated with the integrated intensity of the line integral $W$ (K km s$^{-1}$) \citep{schnee07} as the following,
\begin{equation}
N_{\textrm{thin}} = \frac{8{\pi}W}{{\lambda}^{3}A}\frac{g_{l}}{g_{u}} \frac{1}{J_{\nu}(T_{\textrm{ex}}) - J_{\nu}(T_{\textrm{bg}})}
\frac{1}{1 - \textrm{exp}(-h{\nu}/kT_{\textrm{ex}})} \frac{Q_\textrm{rot}}{g_{l} \textrm{exp}(-E_{l}/kT_{\textrm{ex}})}\:\textrm{cm}^{-2}
\end{equation}
where $\nu$ and $\lambda$ are the frequency and wavelength of the transition, $A$ is the Einstein coefficient, $g_l$ and $g_u$ are the statistical weights of the lower and the upper levels, $J_{\nu}(T) = \frac{h{\nu}/k}{\textrm{e}^{h{\nu}/kT}-1}$, $Q_{\textrm{rot}}$ is the partition function, and $E_l$ is the energy of the lower level.
The energy of the each level is given by $E_{J}=J(J+1)hB$ where $B$ is the rotational constant.

However, in some cases, \ceot\ is not perfectly optically thin.
An optical depth can be corrected in the calculation of column density by using the correction factor, $C_{\tau}$, as long as $\tau \leq 2$ \citep{schnee07}.
\begin{equation}
C_{\tau}=\frac{\tau}{1-e^{-\tau}}.
\end{equation}
The optical depth ($\tau$) can be found from
\begin{equation}
\tau = -\textrm{ln}\left[1-\frac{T_{\textrm{thin}}}{J(T_{\textrm{ex}})-J(T_{\textrm{bg}})}\right],
\end{equation}
where $T_{\textrm{thin}}$ is the beam corrected brightness temperature of the given line.
The final column density corrected for the optical depth is
\begin{equation}
N=N_{\textrm{thin}} \cdot C_{\tau}.
\end{equation}

Finally, the H$_2$ column density is derived by dividing the abundance of the molecule $x$, $X(x)$:
\begin{equation}
N (\textrm{H$_2$}) = \frac{N(x)}{X(x)} \ \textrm{cm}^{-2}
\end{equation}
The C$^{18}$O abundance, $X(\textrm{C$^{18}$O)}$, of $4.8\times 10^{-7}$ is used \citep{lee03}.

The sizes of G108--N and G108--S are defined as $R=\sqrt{a\cdot b}$, where $a$ and $b$ are the major and minor axes of deconvolved FWHM size measured at the 50 $\%$ contours of the \ceot\ emission region.
The mean \atomh\ column densities of G108--N and G108--S within the clump size are 6.7 $\times 10^{21}$ and 1.0 $\times 10^{22}$ cm$^{-2}$, respectively.
Derived parameters are listed in Table 4.

\subsubsection{\atomh\ column density derived from dust continuum emission}

 The molecular gas column density can be calculated with the dust continuum flux, $S_{\nu}$, using
\begin{equation}
N(H_2) = \frac{S_{\nu}}{{\mu}m_{\textrm{H}}{\kappa}_{\nu}B_{\nu}(T_{\textrm{dust}}){\Omega}}\ \textrm{cm}^{-2}
\end{equation}
where $\mu$ is the mean molecular weight, $m_{\textrm{H}}$ is the atomic hydrogen mass, ${\kappa}_{\nu}$ is the mass opacity coefficient, and $B_{\nu}(T_{\textrm{dust}})$ is the Planck function with a dust temperature ($T_{\textrm{dust}}$).
The dust temperature \td\ is adopted from the gas temperature derived in Section 4.2.1.
The aperture solid angle, $\Omega=\pi\theta^2/4 ln2$ is a circular Gaussian aperture where $\theta$ is the beam FWHM.
We use the 850 \m\ dust continuum emission to calculate the \atomh\ column density.
The mass opacity coefficient of 0.018 ${\textrm{cm}}^2 \ {\textrm{g}}^{-1}$ \citep{ossenkopf94} is used, by assuming a gas-to-dust ratio of 100.
In order to make a consistent comparison, the 850 \m\ data have been smoothed to the \ceot\ resolution.
A beam FWHM of 32.5\arcsec\ is used to calculate the aperture solid angle.
Within the clump size, the mean \atomh\ column densities of G108--N and G108--S are 5.6 $\times 10^{21}$ and 8.4 $\times 10^{21}$ cm$^{-2}$, respectively.
Derived parameters are also listed in Table 4.

\subsubsection{Total mass and kinematics}
Dust continuum emission at 850 \m\ is known as one of the best tracers of the total cloud mass because it is optically thin.
Therefore, we derive the total (gas+dust) masses, $M_{\textrm{total}}$, of G108--N and G108--S, using
\begin{equation}
M_{\textrm{total}} = \frac{S_{\nu}{d}^{2}}{{\kappa}_{\nu}B_{\nu}(T_{\textrm{dust}})},
\end{equation}
where S$_{\nu}$ is the 850 \m\ integrated flux and $d$ is the distance \citep[3.21 $\pm$ 0.21 kpc;][]{ramirez11}.
We adopt the same coefficient as those used in Equation (8).
\td\ for G108--N and G108--S are adopted from the gas temperature derived in Section 4.2.1. as 13.4 and 12.5 K, respectively.
Their derived total masses are listed in Table 5.
The total mass of G108--N is larger than the sum of the masses of G108--S1 to S4; the total mass of G108--N is 386 $\pm$ 2 M$_{\sun}$, indicative of a massive starless clump.

To investigate whether the clump is gravitationally unstable, and thus, the possibility of future star formation, we compare its Jeans mass with the total clump mass.
The gas in molecular clouds is supported against gravitational collapse by turbulence and magnetic field as well as the thermal pressure.
The Jeans mass can be derived by
\begin{equation}
M_J \approx \left(\frac{T_{\textrm{eff}}}{10\:\textrm{K}}\right)^{\frac{3}{2}}\left(\frac{\mu}{2.33}\right)^{-\frac{1}{2}}\left(\frac{n}{10^{4}\:\textrm{cm}^{-3}}\right)^{-\frac{1}{2}} M_{\sun},
\end{equation}
taking into account the thermal and turbulent support \citep{hennebelle08}, where $n=N(H_2)/2R$ is the volume density of $H_2$.
The effective kinematic temperature, T$_{\textrm{eff}}$, is given by
\begin{equation}
T_{\textrm{eff}}= C^{2}_{s, \textrm{eff}}  {\mu}m_{\textrm{H}}/ k
\end{equation}
with the effective sound speed C$_{s, {\textrm{eff}}}$.
The C$_{s, {\textrm{eff}}}$ including thermal and turbulent support can be derived by
\begin{equation}
C_{s, {\textrm{eff}}}= \left[({\sigma}_{\textrm{NT}})^{2} + ({\sigma}_{\textrm{T}})^{2}\right]^{\frac{1}{2}}.
\end{equation}
The one-dimensional thermal (${\sigma}_{\textrm{T}}$) and non-thermal (${\sigma}_{\textrm{NT}}$) velocity dispersions can be calculated as follows,
\begin{equation}
{\sigma}_{\textrm{T}}= \left[\frac{kT_{\textrm{ex}}}{m_{\textrm{H}}\mu}\right]^{\frac{1}{2}}
\end{equation}
\begin{equation}
{\sigma}_{\textrm{NT}}= \left[{\sigma}^{2}_{\textrm{C$^{18}$O}} - \frac{kT_{\textrm{ex}}}{m_{\textrm{c$^{18}$O}}}\right]^{\frac{1}{2}},
\end{equation}
where $k$ is the Boltzmann's constant.
${\sigma}^{2}_{\textrm{C$^{18}$O}}=\Delta V^{2}_{\textrm{C$^{18}$O}}/8ln(2)$ is the one-dimensional velocity dispersion of \ceot\ and $m_{\textrm{C$^{18}$O}}$ is the mass of C$^{18}$O.
The calculated Jeans mass is 299 $M_{\sun}$ with $n=1.0\times10^3$ cm$^{-3}$, $T_{\textrm{eff}}=222$ K, $\sigma_{\textrm{T}}=0.20$ km s$^{-1}$, and $\sigma_{\textrm{NT}}=0.79$ km s$^{-1}$.
The total mass of G108--N (386 $M_{\sun}$) is larger than $M_J$, suggesting that G108--N is gravitationally unstable (i.e., a prestellar clump).

\subsection{Chemical status of two clumps}

\subsubsection{CO depletion}
According to the comparison between the integrated intensity distributions of molecular lines and dust continuum, the molecular line emission peaks are slightly shifted to SW from the dust continuum emission peak.
This might be caused by the molecular depletion at the dust continuum peak \citep{lee03}.
We compare the H$_{2}$ column densities, \cd, derived from \ceot\ and the 850 \m\ dust continuum emission to derive the CO depletion factor, which is defined as:
\begin{equation}
D_{\textrm{co}}= \frac{N({\textrm{H}}_2)_{\textrm{dust}}}{N({\textrm{H}}_2)_{{\textrm{C}}^{18}{\textrm{O}}}}.
\end{equation}

The maximum depletion factor of G108--N is about 1.8 while that of G108--S is about 4.
The low CO depletion suggests that the dynamical timescales of G108--N and G108--S are not very large, that is, the high-density clumps might form recently.
However, we cannot rule out an artificial effect; the \atomh\ column density derived from dust continuum emission could be underestimated because the extended structure is possibly filtered out during the reduction process.

\subsubsection{Molecular abundances}

Figure 12 shows the averaged spectra within the 50 $\%$ contour of integrated intensity peak of \nhp\ towards G108--N and G108--S obtained with the NRO 45 m telescope; the \nhp\ emission is spatially distributed in larger area than the beam FWHM (20.4 $\arcsec$).
The \nhp\ line is fitted using the hyperfine structure (HFS) fitting method in GILDAS/CLASS to determine \tex\ and optical depth, $\tau$.
In both clumps, the \nhp\ lines are optically thin ($\tau < 0.1$).
Derived \tex\ of G108--N and G108--S are 3.7 and 3.3 K, respectively, which are very low, compared to the \tex\ of CO in G108--N (13.4 K) and G108--S (12.5 K).
We also derived the \tex\ of HCN by fitting the HFS of the \hcn\ line.
The spectra of \hcn\ at each positions are presented in Figure 13.
The \tex\ of HCN in G108--N and G108--S are 3.6 and 3.8 K, respectively, which are similar to those of N$_{2}$H$^{+}$.
The optical depths at all five positions are less than 1.

N$_{2}$H$^{+}$ and HCN seem sub-thermally excited, so we derive column density of N$_{2}$H$^{+}$ and HCN using a non-LTE radiative transfer code, RADEX \citep{vandertak07}.
We assume that \tkin\ is equal to \tex\ derived from \ceot\ for G108--N and G108--S.
The mean \atomh\ volume density is derived from the 850 \m\ dust continuum emission.
For the N$_{2}$H$^{+}$ analysis, the 850 \m\ dust continuum map is convolved with the beam FWHM of NRO 45 m.
The line widths of isolated component of \nhp\ ($F_{1}, F=0,1 \rightarrow 1,2$) of 1.72 and 2.67 km s$^{-1}$ for G108--N and G108--S, respectively, are used.
Calculated \nhp\ column densities of G108--N and G108--S are 1.2 $\times 10^{14}$ and 1.9 $\times 10^{14}$ cm$^{-2}$, respectively.
\tex\ from non-LTE calculation are 3.1 and 3.0 K for G108--N and G108--S, respectively, which are consistent with the result from the hyperfine structure (HFS) fitting method.
The optical depth, however, calculated using non-LTE method are larger than that calculated by the HFS fitting.
Fractional abundance of molecule $x$, $X(x)$, can be derived as follows,
\begin{equation}
X(x) = N(x) / N(\textrm{H}_{2})_\textrm{dust}.
\end{equation}
The fractional abundances of \nhp\ are 1.2 $\times 10^{-8}$ and 2.5 $\times 10^{-8}$ for G108--N and G108--S, respectively.
G108--N and G108--S have higher fractional abundances than previously studied dense cores ($X(\textrm{N}_{2}\textrm{H}^{+}) \sim 1 \times 10^{-10}$) by \citet{caselli02}, \citet{tafalla02}, and \citet{di04}.
On the other hand, the \nhp\ fractional abundances of two clumps are consistent with those of cores within ``Clump--S" of PGCC G192.32--11.88 \citep{liu16}, which are calculated in the assumption of LTE.
The \tex\ of cores in ``Clump-S" is about 5 K and the optical depths is about 4.
Derived parameters using \nhp\ are listed in Table 6.

For the non-LTE analysis of \hcn, the 850 \m\ dust continuum map is convolved with the beam FWHM of KVN 21 m to derive the \atomh\ volume desity.
The \hcn\ spectra at the positions of the 850 \m\ emission peaks are presented in Figure 11.
The line width of the strongest component of \hcn\ is used.
We average the four values for G108--S1, S2, S3, and S4 as a representative value for G108-S for the consistency with the N$_{2}$H$^{+}$ analysis.
\tex\ from the non-LTE calculation for \hcn\ are 3.8 and 4.0 K toward G108--N and G108--S, respectively.
The column densities of \hcn\ in G108--N and G108--S are 7.8 $\times 10^{14}$ and 6.5 $\times 10^{14}$ cm$^{-2}$, respectively.
The optical depth from the non-LTE method are larger than that from the HFS analysis.
The fractional abundances of \hcn\ are 3.9 $\times 10^{-7}$ and 2.9 $\times 10^{-7}$ for G108--N and G108--S, respectively.
Derived parameters with \hcn\ are listed in Table 7.

\section{DISCUSSIONS}

\sou\ is located close to dynamically active regions such as an H{\sc ii} region (Sh2-152) and a SNR (SNR G109.1--1.0).
SNR G109.1--1.0 is located at the east of \sou.
A line profile composed of two components, a sharp single peaked component with a broad component, was shown in the interacting regions between CO molecular clouds and SNRs \citep{seta98,su14}.
So, the association between SNR G109.1--1.0 and CO molecular cloud in the vicinity of the remnant has been previously investigated \citep{tatematsu87,tatematsu90}.
However, according to \citet{tatematsu90}, any broad CO emission, which is possibly accelerated by SNR G109.1--1.0, was not detected.

The channel map of \tco\ line emission of \sou\ are shown in Figure 14. 
The CO emission distributes along two filaments. 
A thin filament to the east is seen from --55 to --52 km~s$^{-1}$. 
The main filament, which includes G108--N and G108--S, is seen from --52 to --46.5 km~s$^{-1}$. 
The two filaments may be interacting in the north-east region, where IRAS 22576+5843 is located. 
However, the broad CO emission, which could be a hint for the interaction, does not appear. 
The thin filament is located in front of the main filament and seems to make a shield against the SNR shocks. 
As mentioned in section 3.2, \tcot$/$\tco\ ratios along the main filament are not enhanced by shocks. 
Therefore, the main filament seems unaffected by the SNR.

However, the star formation in G108--S seems affected by the southern H{\sc ii} region, Sh2-152; G180--S is bent probably due
to external compression by Sh2-152 and the \tcot$/$\tco\ ratio is enhanced at the bent region (see Figure 5). 
The gravitational collapse and fragmentation of G108--S might be induced by Sh2-152.
G108--S is more evolved than G108--N.
G108--N is a massive prestellar clump while G108--S is fragmented into 4 cores  at least , G108--S1, G108--S2, G108--S3, and G108--S4, and G108--S1 (IRS 1) and G108--S2 (IRS 2) are associated with Class I sources.
The overall clump properties of G108--N and G108--S are similar, and their chemical status does not show significant difference; G108--S is slightly more evolved than G108--N in the view of CO depletion.

By comparing the pressures of G108--S and Sh2-152, one can investigate whether the evolution of G108--S has been affected by Sh2-152.
The internal pressure of Sh2-152 can be derived using \citep{morgan04}
\begin{equation}
\frac{P_{\textrm{H{\sc ii}}}}{k}= 2n_{\textrm{e}}T_{\textrm{e}}.
\end{equation}
The effective electron temperature $T_\textrm{e}$ is assumed to be $10^4$ K.
For the electron density, the peak value ($n_\textrm{e}=2000$ cm$^{-3}$) from \citet{heydari81} is used. 
The derived $P_{\textrm{H{\sc ii}}}/k$ is $4.0\times 10^7$ cm$^{-3}$ K.
On the other hand, the molecular pressure inside G108--S is \citep{liu12a}
\begin{equation}
\frac{P_{\textrm{mol}}}{k}=nT_{\textrm{eff}}.
\end{equation}
The $P_{\textrm{mol}}/k$ is derived to be $7.3\times 10^5$ cm$^{-3}$ K with $n=4.5\times10^3$ cm$^{-3}$ and $T_\textrm{eff}=83$ K, which are obtained from the \ceot\ emission.
The pressure of Sh2-152 is two orders of magnitude larger than the internal pressure of G108--S.
It suggests that G108--S could be compressed by Sh2-152 if the two are physically associated.
More studies are needed to confirm whether the star formation in G108--S has been indeed induced by Sh2-152. 
In contrast, G108--N is quiescent and seems unaffected by its surrounding environment yet.

The stability of G108--N was examined by comparing its total mass and the Jeans mass; the derived total mass is greater than the Jeans mass, indicating that G108--N is gravitationally unstable. 
The free-fall time \citep[$t_{\textrm{ff}}=3.4\times10^7 n^{-0.5}$ yr,][]{evans99} for G108--N is about 1 Myr, which is similar to the typical lifetime of low mass starless cores with the average density of $n\sim10^{4}$ cm$^{-3}$.
The lifetime for massive prestellar cores is expected to be shorter than the free-fall time \citep{andre14}.
Therefore, G108-N might be about to collapse.
However, it is unclear whether G108--N will survive as a single massive prestellar core or be fragmented into several components as turned out in G108--S.
The peak surface densities of G108--N and G108--S are similar to 0.07 g cm$^{-2}$ as derived from the \ceo\ line.
The value is much smaller than the threshold surface density \citep[1 g cm$^{-2}$,][]{krumholz08} for fragmentation in massive star formation.
The estimated surface densities of IRDCs are 0.2 to 5 g cm$^{-2}$ \citep{battersby11}.
However, \citet{urquhart14} argued that adding the effect of magnetic field could reduce the threshold of surface density for fragmentation.
In addition, the distribution of surface densities of massive star-forming clumps shows that clumps with the surface density lower than 0.1 g cm$^{-2}$ can result in high mass star formation \citep[Figure 12 of][]{urquhart14}. 
Therefore, we cannot exclude the possibility that G108--N could form a high-mass star without fragmentation.

\section{SUMMARY}

To investigate the star formation condition near active regions such as SNRs and H{\sc ii} regions, various molecular line observations as well as the submm dust continuum observation toward \sou\ were carried out.
Our results are:

1. One filament structure presented in the integrated intensity maps of \tco\ and \thco\ is largely divided into three parts: the northeast region associated with a star, IRAS 22576+5843, the southern region directly linked to H{\sc ii} region, Sh2-152, and the central part defined here as G108 containing IRAS 22565+5839.
The two clumps, ``G108--N'' and ``G108--S'' are found within G108.

2. In G108--S, a blue-shifted component, whose velocity is similar to that of the component associated with Sh2-152, 
is shown in both of the Moment 1 maps of \tcot\ and \thcot.
The integrated intensity ratio, \thcot$/$\thco, is also the most enhanced at the southern edge of G108--S, which is close to the H{\sc ii} region.

3. Dust continuum emission also shows two components corresponding to G108--N and G108--S.
Highly fragmented structures in G108--S are revealed with the beam FWHM of JCMT/SCUBA-2 ($\sim14\arcsec$).
Four components, G108--S1, G108--S2, G108--S3, and G108--S4, are identified in the G108--S.
G108--N has no associated IR source, while there are two IR sources corresponding with the dust continuum emission peaks, G108--S1 and G108--S2.
The integrated intensity peaks of \tcot, \thcot\, and \ceot\ are slightly off that of dust continuum.

4. The \hcop\ and \hcn\ show similar distributions to each other and also similar to the distribution of 850 \m\ continuum emission
in G108--N.
The emission of two lines is distributed along the NE--SW direction with a cometary structure while \hcopf\ is shown only in the head part.
This cometary structure is also shown in the integrated intensity map of \nht.
The centroid velocities of all lines are consistent.

5. IR sources, IRS 1 and IRS 2, associated with G108--S1 and G108--S2, respectively, are YSO candidates.
IRS 1 and IRS 2 are classified as Class I sources.
However, outflows associated with these Class I sources were not detected with our CO observations.
Higher resolutions and better sensitivities might reveal associated outflows.

6. The \tex\ derived from CO observations of G108--N and G108--S are 13.4 and 12.5 K, respectively.
Total mass of G108--N is larger than the sum of total masses of G108 S1 to S4.
The total mass of G108--N is larger than the Jeans mass, suggesting that G108--N is a prestellar clump, which is gravitationally unstable and thus a potential future star formation site.

7. Two clumps have similar properties such as the gas temperature and the \atomh\ column density.
Their chemical status does not show significant difference either although G108--N is slightly less evolved than G108--S in the view of CO depletion.

8. The gravitational fragmentation and star formation in G108-S seem induced by the compression from the H{\sc ii}, Sh2-152; G108--S is bent toward Sh2-152 and the \tcot$/$\tco\ ratio is enhanced at the bent region.  
In contrast, G108--N does not seem affected by its environment yet.

This work was supported by the Basic Science Research Program through the National Research Foundation of Korea (NRF) (grant No.NRF-2015R1A2A2A01004769) and the Korea Astronomy and Space Science Institute under the R\&D program (Project No. 2015-1-320-18) supervised by the Ministry of Science, ICT and Future Planning.
Tie Liu is supported by the KASI and EACOA fellowships. The James Clerk Maxwell Telescope is operated by the East Asian Observatory on behalf of The National Astronomical Observatory of Japan, Academia Sinica Institute of Astronomy and Astrophysics, the Korea Astronomy and Space Science Institute, the National Astronomical Observatories of China and the Chinese Academy of Sciences (Grant No. XDB09000000), with additional funding support from the Science and Technology Facilities Council of the United Kingdom and participating universities in the United Kingdom and Canada.
The KVN is a facility operated by the Korea Astronomy and Space Science Institute.

\begin{deluxetable}{lcccc}
\tabletypesize{\footnotesize}
\tablecaption{Summary of the Observations}
\tablewidth{0pt}
\tablehead{	\colhead{} &
			\colhead{$\nu$} &
			\colhead{Beam FWHM} &
			\colhead{$\eta_{mb}$\tablenotemark{a}} &
			\colhead{Telescope} \\
			\colhead{Molecular line} &
			\colhead{[GHz]} &
			\colhead{[arcsec]} &
			\colhead{} &
			\colhead{}}

\startdata
{\hcn} & 88.631847 & 29 & 0.81 & IRAM 30 m\\
           &                    & 32 & 0.37 & KVN 21 m (Yonsei) \\
           &                    & 33 & 0.43 & KVN 21 m (Ulsan) \\
           &                    & 32 & 0.36 & KVN 21 m (Tamna) \\
\hcop  & 89.188526 & 29 & 0.81 &  IRAM 30 m\\
\nhp  & 93.173776 & 20.4 & 0.54 &  NRO 45 m\\\
\tco & 115.271202 & 60 & 0.67 & PMO 13.7 m\\
\thco &110.201354& 60 & 0.67 &  PMO 13.7 m\\
\ceo & 109.782176 & 60 & 0.67 & PMO 13.7 m\\
\tcot & 230.538000 & 32.5 & 0.76 & CSO 10 m\\
\thcot & 220.398684 & 32.5 & 0.76 & CSO 10 m\\
\ceot & 219.560358 & 32.5 & 0.76 &  CSO 10 m\\
\nht & 236.94495 & 40 & 0.79 & Effelsberg 100 m

\enddata
\tablenotetext{a}{\textbf{The main beam efficiencies}}
\end{deluxetable}
\clearpage

\begin{deluxetable}{ccccccccccc}
\tabletypesize{\tiny}
\tablewidth{0pt}
\rotate
\tablecaption{Infrared Fluxes of YSO Candidates from the Archive}
\tablehead{ \colhead{} &
			\colhead{RA} &
			\colhead{Dec} &
			\colhead{2MASS H} &
			\colhead{2MASS K$_\textrm{s}$} &
			\colhead{WISE 3.4 $\mu$m} &
			\colhead{WISE 4.6 $\mu$m} &
			\colhead{AKARI 9 $\mu$m} &
			\colhead{WISE 12 $\mu$m} &
			\colhead{AKARI 18 $\mu$m} &
			\colhead{WISE 22 $\mu$m} \\
			\colhead{Source} &
			\colhead{($^{h}$ $^{m}$ $^{s}$)} &
			\colhead{($\degr$ $\arcmin$ $\arcsec$)} &
			\colhead{(mJy)} &
			\colhead{(mJy)} &
			\colhead{(mJy)} &
			\colhead{(mJy)} &
			\colhead{(mJy)} &
			\colhead{(mJy)} &
			\colhead{(mJy)} &
			\colhead{(mJy)} }
\startdata
IRS 1 & 22:58:34.08 & +58:55:51.88 & \nodata & \nodata & 14.9 (0.2) & 28.5 (0.5) & 118.0 (18.6) & 83.9 (1.8) & 320.0 (21.0) & 291.0 (6.0) \\
IRS 2 & 22:58:38.56 & +58:55:48.39 & 2.13 (0.16) & 11.1 (0.5) & 41.4 (0.8) & 68.9 (1.3) & 107 (3.0) & 88.5 (1.5) & \nodata & 196 (4.0)  \\

\enddata
\end{deluxetable}
\clearpage

\begin{deluxetable}{cccc}
\tablecaption{Bolometric Luminosities, Bolometric Temperatures, and IR Spectral Indices of IRS 1 and IRS 2}
\tablewidth{0pt}
\tablehead{        \colhead{Sources} &
			\colhead{$L_{\rm{bol}}$} &
			\colhead{$T_{\rm{bol}}$ (Class) } &
			\colhead{Spectral index $\alpha$ (Class)}\\
			\colhead{} &
			\colhead{[$L_{\sun}$]} &
			\colhead{[K]} &
			\colhead{}}		
\startdata
IRS 1 & 27  & 341 (I)  & 0.61 (I)  \\
IRS 2 & 30  & 614 (I) & 0.48 (I) \\
\enddata

\end{deluxetable}
\clearpage

\begin{deluxetable}{cccccccc}
\tablewidth{0pt}
%\rotate
\tabletypesize{\small}
\tablecaption{Derived parameters of \ceo\ of G108--N and G108--S from LTE analysis }
\tablehead{\colhead{} &
        \colhead{} &
        \colhead{peak} &
        \colhead{mean} &
        \colhead{peak} &
        \colhead{mean} &
        \colhead{} &
        \colhead{} \\
		\colhead{ID} &
 		\colhead{$T_{\textrm{ex}}$} &
        \colhead{$N(\textrm{H}_2)_{\textrm{C}^{18}\textrm{O}}$} &
        \colhead{$N(\textrm{H}_2)_{\textrm{C}^{18}\textrm{O}}$} &                  	
        \colhead{$N(\textrm{H}_2)_{\textrm{dust}}$} &
        \colhead{$N(\textrm{H}_2)_{\textrm{dust}}$} &
        \colhead{a\tablenotemark{a}} &
        \colhead{b\tablenotemark{b}} \\
        \colhead{} &
        \colhead{(K)} &
        \colhead{($\times10^{21}$ cm$^{-2}$)} &
        \colhead{($\times10^{21}$ cm$^{-2}$)} &
        \colhead{($\times10^{21}$ cm$^{-2}$)} &
        \colhead{($\times10^{21}$ cm$^{-2}$)} &
        \colhead{(arcsec)} &
        \colhead{(arcsec)}}

\startdata
G108--N & 13.4 &  16.1 & 6.7 & 12.3 & 5.6 & 114 & 40  \\
G108--S & 12.5 & 15.3 & 10.0 & 10.4 & 8.4 & 24 & 16
\enddata
\tablenotetext{a}{\textbf{The size of semi-major axis}}
\tablenotetext{b}{\textbf{The size of semi-minor axis}}
\end{deluxetable}

\begin{deluxetable}{cccc}
\tabletypesize{\footnotesize}
\tablecaption{Total Masses of 850 \m\ Dust Continuum Condensations in G108 }
\tablewidth{0pt}
\tablehead{ \colhead{ID} &
			\colhead{ Integrated Flux Density} &
			\colhead{ Mass } \\
			\colhead{} &
			\colhead{(mJy)} &
			\colhead{($M_{\sun}$})}
			
\startdata
G108--N & 3584.0 & 386\\
G108--S1 & 418.2 & 51 \\
G108--S2 & 373.3 & 45 \\
G108--S3 & 510.1 & 62 \\
G108--S4 & 409.5 & 50 \\

\enddata

\end{deluxetable}

\clearpage

\begin{deluxetable}{cccccc}
\tablewidth{0pt}
\tablecaption{Derived Parameters of \nhp\ from Non-LTE Calculation}
\tablehead{\colhead{ID} &
 		   \colhead{\tkin} &
           \colhead{$N(\textrm{H}_2)_{\textrm{dust}}$} &
           \colhead{$\Delta v$} &
           \colhead{$N(\textrm{N}_2\textrm{H}^+)$} &
           \colhead{$X(\textrm{N}_2\textrm{H}^+)$} \\
           \colhead{} &
           \colhead{(K)} &
           \colhead{($\times10^{21}$ cm$^{-2}$)} &
           \colhead{(km s$^{-1}$)} &
           \colhead{($\times10^{14}$ cm$^{-2}$)} &
           \colhead{($\times10^{-8}$)}}

\startdata
G108--N & 13.4 & 9.9 & 1.72 & 1.2 & 1.2 \\
G108--S & 12.5 & 7.8 & 2.67 & 1.9 & 2.5
\enddata
\end{deluxetable}

\begin{deluxetable}{cccccc}
\tablewidth{0pt}
\tablecaption{Derived parameters of \hcn\ from non-LTE calculation}
\tablehead{\colhead{ID} &
 		   \colhead{\tkin} &
           \colhead{$N(\textrm{H}_2)_{\textrm{dust}}$} &
           \colhead{$\Delta v$} &
           \colhead{$N(\textrm{HCN})$} &
           \colhead{$X(\textrm{HCN})$} \\
           \colhead{} &
           \colhead{(K)} &
           \colhead{($\times10^{21}$ cm$^{-2}$)} &
           \colhead{(km s$^{-1}$)} &
           \colhead{($\times10^{14}$ cm$^{-2}$)} &
           \colhead{($\times10^{-7}$)}}

\startdata
G108--N & 13.4 & 7.9 & 3.24 & 7.8 & 3.3 \\
G108--S & 12.5 & 7.4 & 2.53 & 6.5 & 2.9
\enddata
\end{deluxetable}

\clearpage

\begin{figure}[!p]
\begin{center}
\plotone{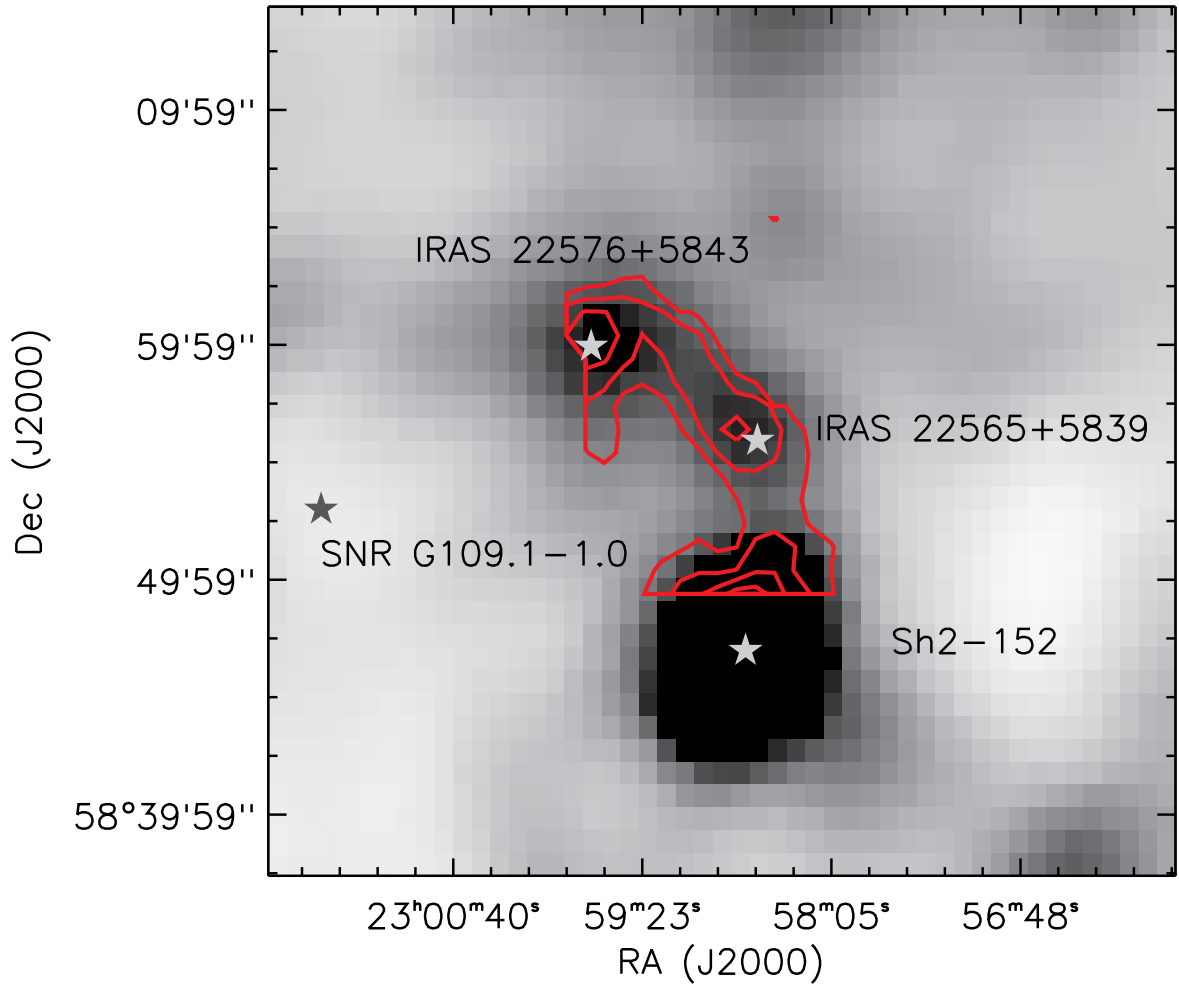}
\caption{Integrated intensity maps of \tco\ (red contours) overlaid on \textit{Planck} 350 \m\ image (gray scale).
Contour intervals are 20 $\%$ of the each peak intensity and range from 30 $\%$ to 90 $\%$.
The peak integrated intensity of \tco\ is 77.5 K km s$^{-1}$.}
\end{center}
\end{figure}
\clearpage

\begin{figure}[!p]
\begin{center}
\plotone{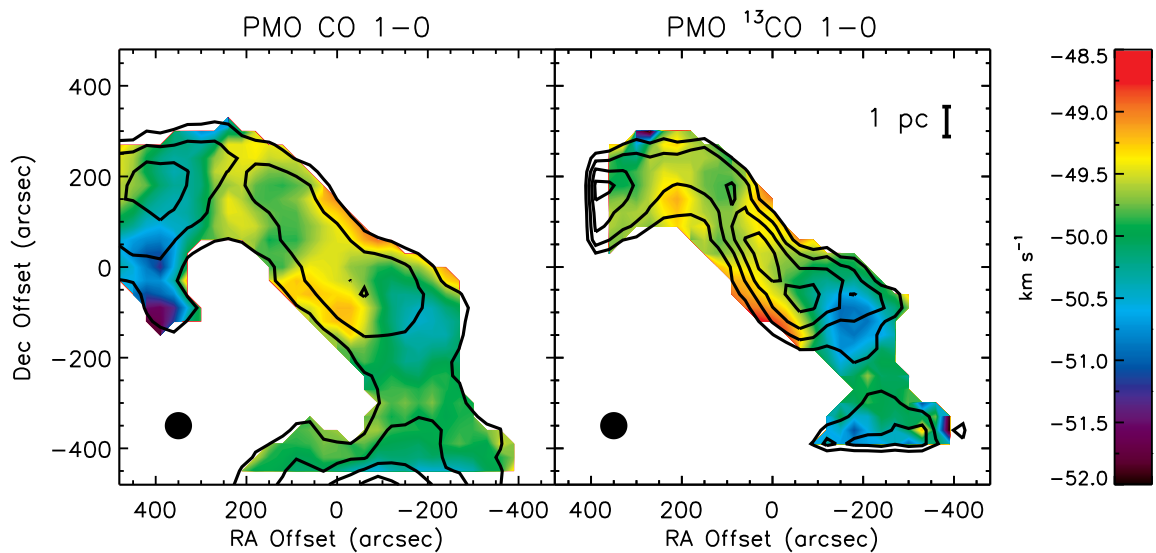}
\caption{Integrated intensity maps (contours) of \tco\ (\textit{left}) and \thco\ (\textit{right}) on top of the Moment 1 maps of \tco\ and \thco\ (color scale).
Contour intervals are 20 $\%$ of the each peak integrated intensity and range from 30 $\%$ to 90 $\%$.
The peak integrated intensities of \tco\ and \thco\ are 77.5 and 19.5 K km s$^{-1}$, respectively.
The Moment 1 maps are constructed with the data clipped using 3$\sigma$ criteria.
The filled circle at the bottom-left corner of each panel denotes the respective beam FWHM of PMO 14 m.}
\end{center}
\end{figure}
\clearpage

\begin{figure}[!p]
\begin{center}
\plotone{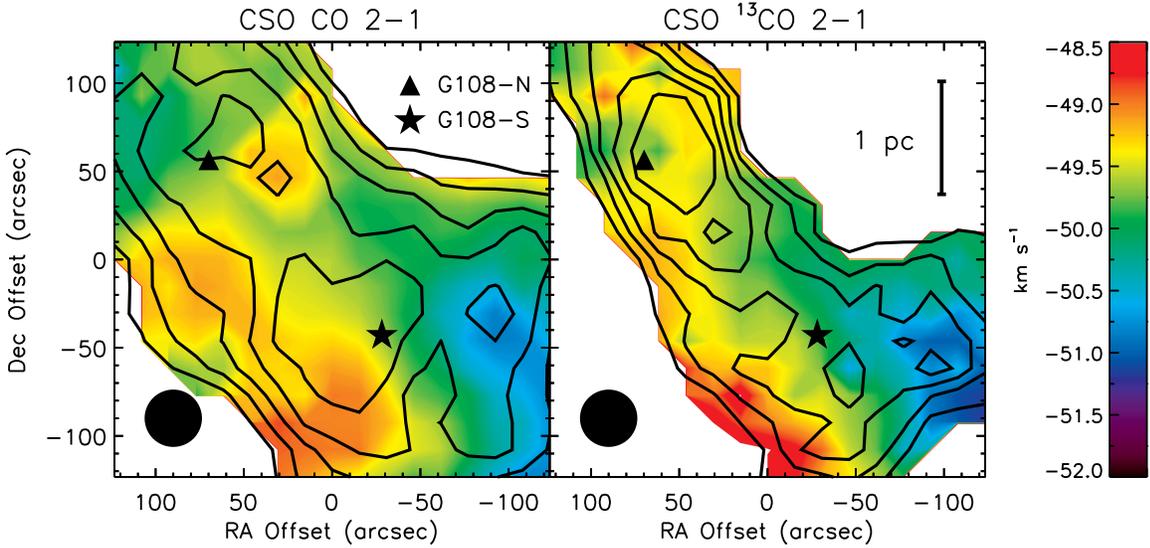}
\caption{Integrated intensity maps (contours) of \tcot\ (\textit{left}) and \thcot\ (\textit{right}) on top of the Moment 1 maps of \tcot\ and \thcot\ (color scale).
Contour intervals are 10 $\%$ of the each peak intensity and range from 50 $\%$ to 90 $\%$.
The peak integrated intensities of \tcot\ and \thcot\ are 50.9 and 18.5 K km s$^{-1}$, respectively.
The Moment 1 maps are constructed with the data clipped using 3$\sigma$ criteria.
The filled circle at the bottom-left corner of each panel denotes the respective beam FWHM of CSO.
The positions of G108--N and G108--S are denoted by a filled triangle and star, respectively.}
\end{center}
\end{figure}
\clearpage

\begin{figure}[!p]
\epsscale{0.7}\begin{center}
\plotone{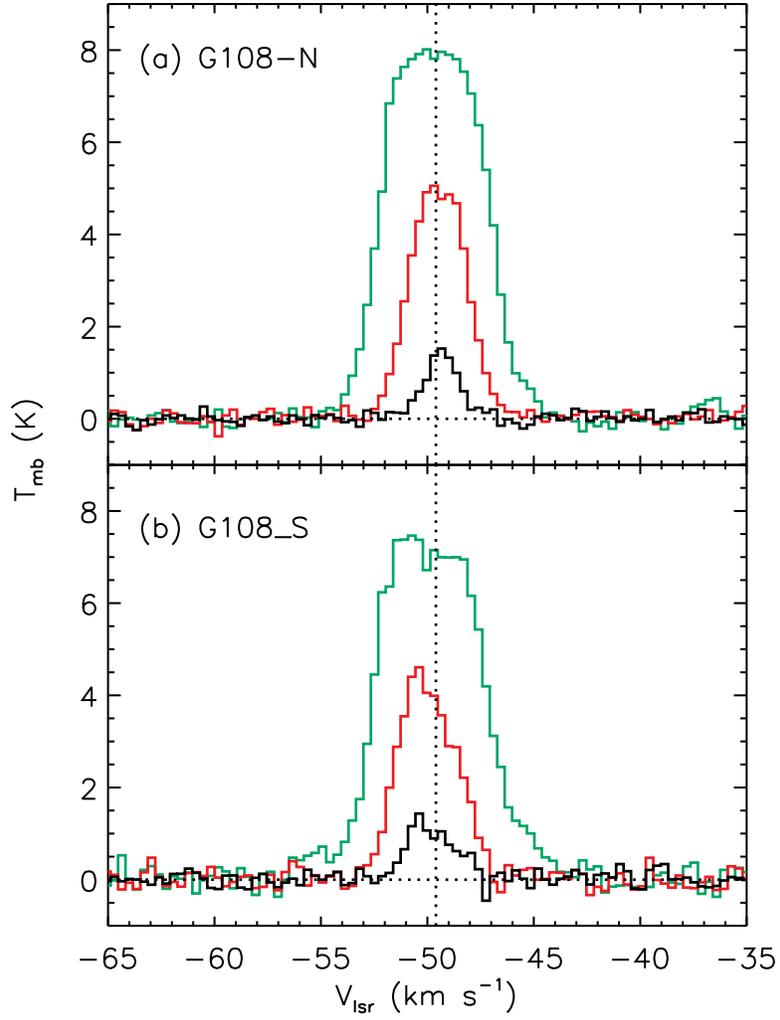}
\caption{Spectra toward G108--N (\textit{left}) and G108--S (\textit{right}).
Spectra of \tcot, \thcot, and \ceot\ are presented in green, red, and black, respectively.}
\end{center}
\end{figure}

\clearpage

\begin{figure}[!p]
\begin{center}
\epsscale{1.0}
\plotone{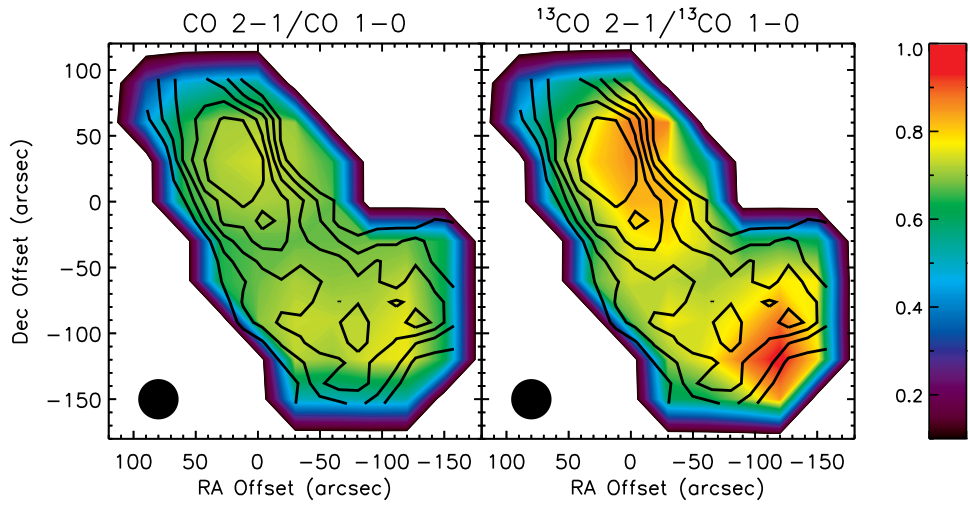}
\caption{Color images show the integrated intensity ratios of \tcot\ to \tco\ (\textit{left}) and \thcot\ to \thco\ (\textit{right}).
The black contours represent integrated intensity of \thcot.
The contours are the same levels as in Figure 2.
}
\end{center}
\end{figure}
\clearpage

\begin{figure}[!p]
\begin{center}
\epsscale{1.0}
\plotone{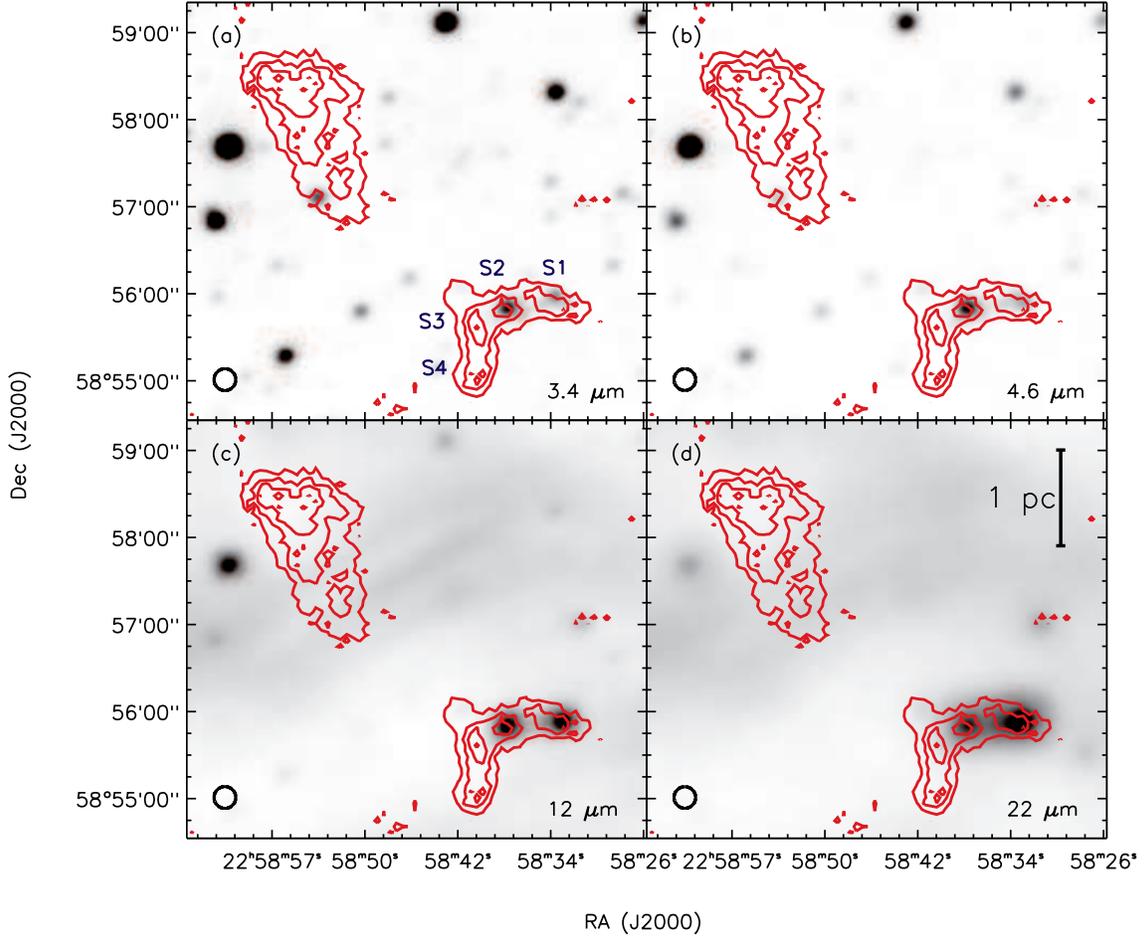}
\caption{Continuum emission at 850 \m\ (red contours) overlaid on the WISE images (gray scale) at (a) 3.4 \m, (b) 4.6 \m, (c) 12 \m, and (d) 22 \m.
Contour levels are 30, 50, 70, and 90 $\%$ of the peak intensity of 232 mJy beam$^{-1}$.
The beam FWHM of JCMT/SCUBA--2 at 850 \m\ is presented with open circle.
}
\end{center}
\end{figure}
\clearpage

\clearpage

\begin{figure}[!p]
\begin{center}
\epsscale{1}

\plotone{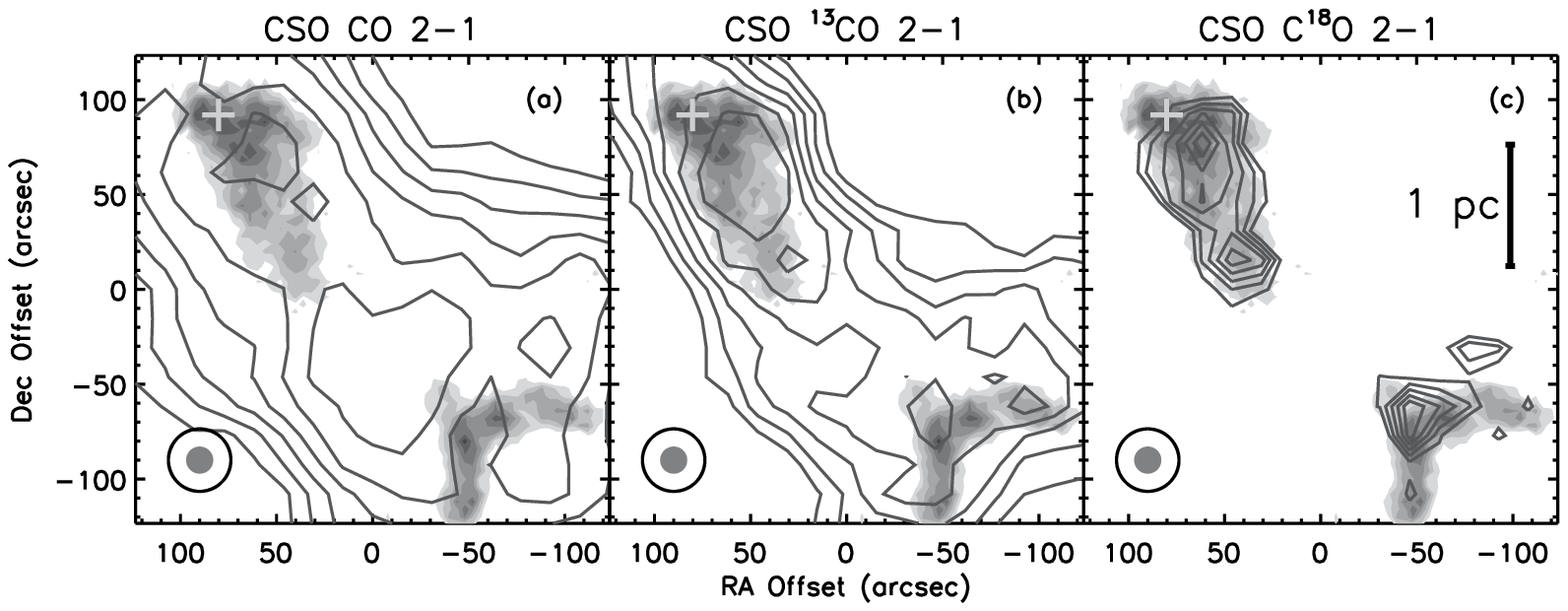}
\epsscale{1}
\caption{Integrated intensity maps (contours) of \tcot\ (\textit{left}), \thcot\ (\textit{center}), \ceot\ (\textit{right}) on top of the 850 \m\ dust continuum map (gray scale).
Contour intervals are 10 $\%$ of the each peak intensity and range from 40 $\%$ to 90 $\%$.
The gray scale levels are 20 to 90 $\%$ of the peak intensity in step of 10 $\%$.
The peak intensities of \tcot, \thcot, and \ceot\ are 50.9, 18.5, and 4.0 K km s$^{-1}$, respectively, while the peak intensity of 850 \m\ dust continuum is 232 mJy beam$^{-1}$.
The filled circle and the opened circle at the bottom-left corner of each panel denotes the respective beam FWHM of JCMT/SCUBA--2 and that of CSO 10 m, respectively.}
\end{center}
\end{figure}
\clearpage

\begin{figure}[!p]
\begin{center}
\epsscale{1}
\plotone{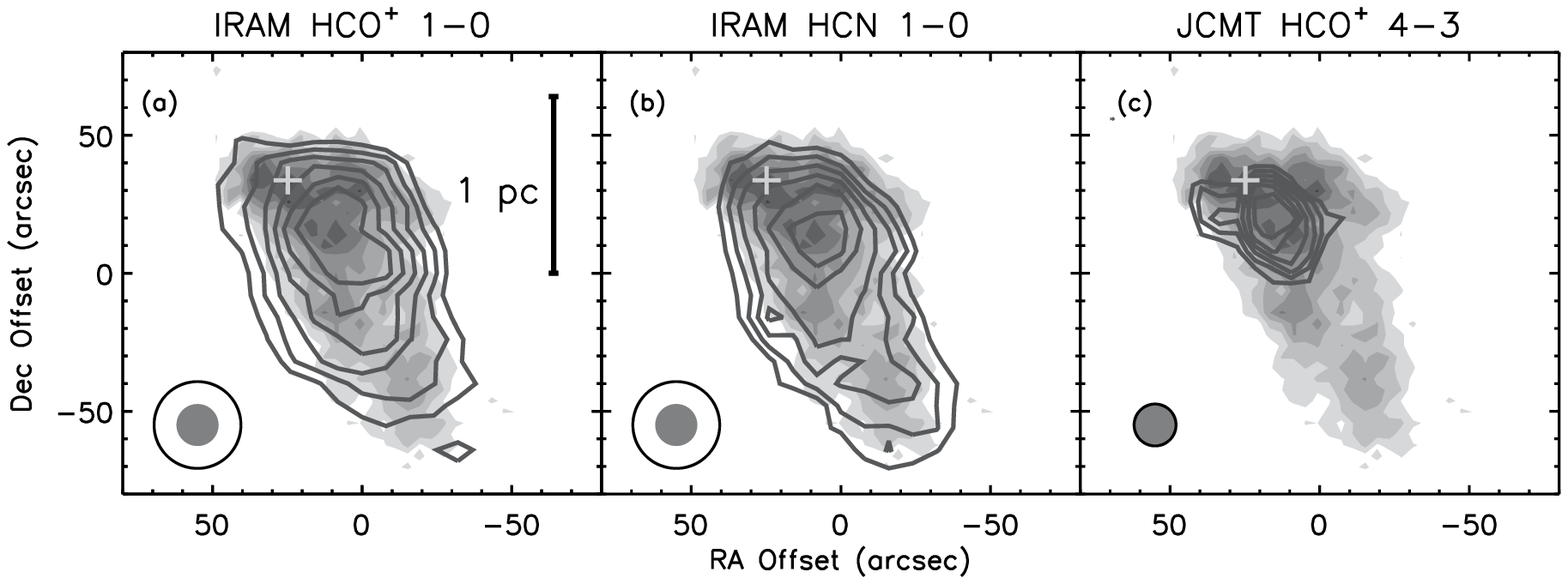}
\caption{Integrated intensity maps (contours) of (a) \hcop, (b) \hcn, and (c) \hcopf\ on top of the 850 \m\ dust continuum map of G108--N (gray scale).
Contour intervals are 10 $\%$ of the each peak intensity and range from 40 $\%$ to 90 $\%$.
The gray scale levels are 20 to 90 $\%$ of the peak intensity in step of 10 $\%$.
The peak intensities of \hcop, \hcn, and \hcopf\ are 4.7, 5.6, and 0.8 K km s$^{-1}$, respectively, while the peak intensity of 850 \m\ dust continuum is 232 mJy beam$^{-1}$.
The filled circle and the opened circle at the bottom-left of each panel denotes the respective beam FWHM of JCMT/SCUBA--2 and that of IRAM 30 m, respectively.}
\end{center}
\end{figure}
\clearpage

\begin{figure}[!p]
\begin{center}
\epsscale{0.5}
\plotone{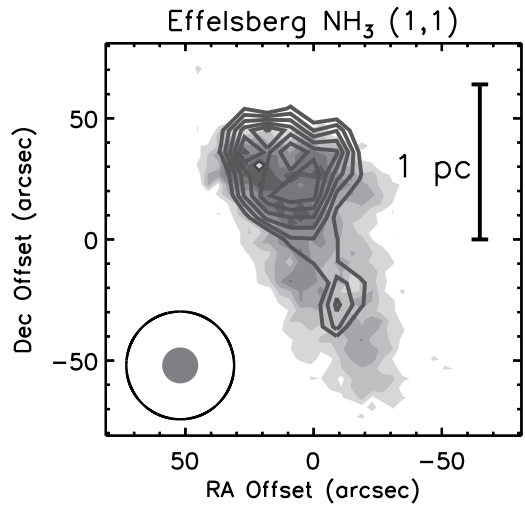}
\caption{Integrated intensity map of \nht\ (contours) on top of the 850 \m\ map of G108--N (gray scale).
Contour intervals are 10 $\%$ of the each peak intensity and range from 40 $\%$ to 90 $\%$.
The gray scale levels are 20 to 90 $\%$ of the peak intensity in step of 10 $\%$.
The peak integrated intensity of \nht\ is 0.54 K km s$^{-1}$, respectively, while the peak intensity of 850 \m\ dust continuum is 232 mJy beam$^{-1}$.
The filled circle and the opened circle at the bottom-left of each panel denotes the respective beam FWHM of JCMT/SCUBA--2 and that of Effelsberg 100 m, respectively.}
\end{center}
\end{figure}
\clearpage

\begin{figure}[!p]
\epsscale{0.4}
\begin{center}
\plotone{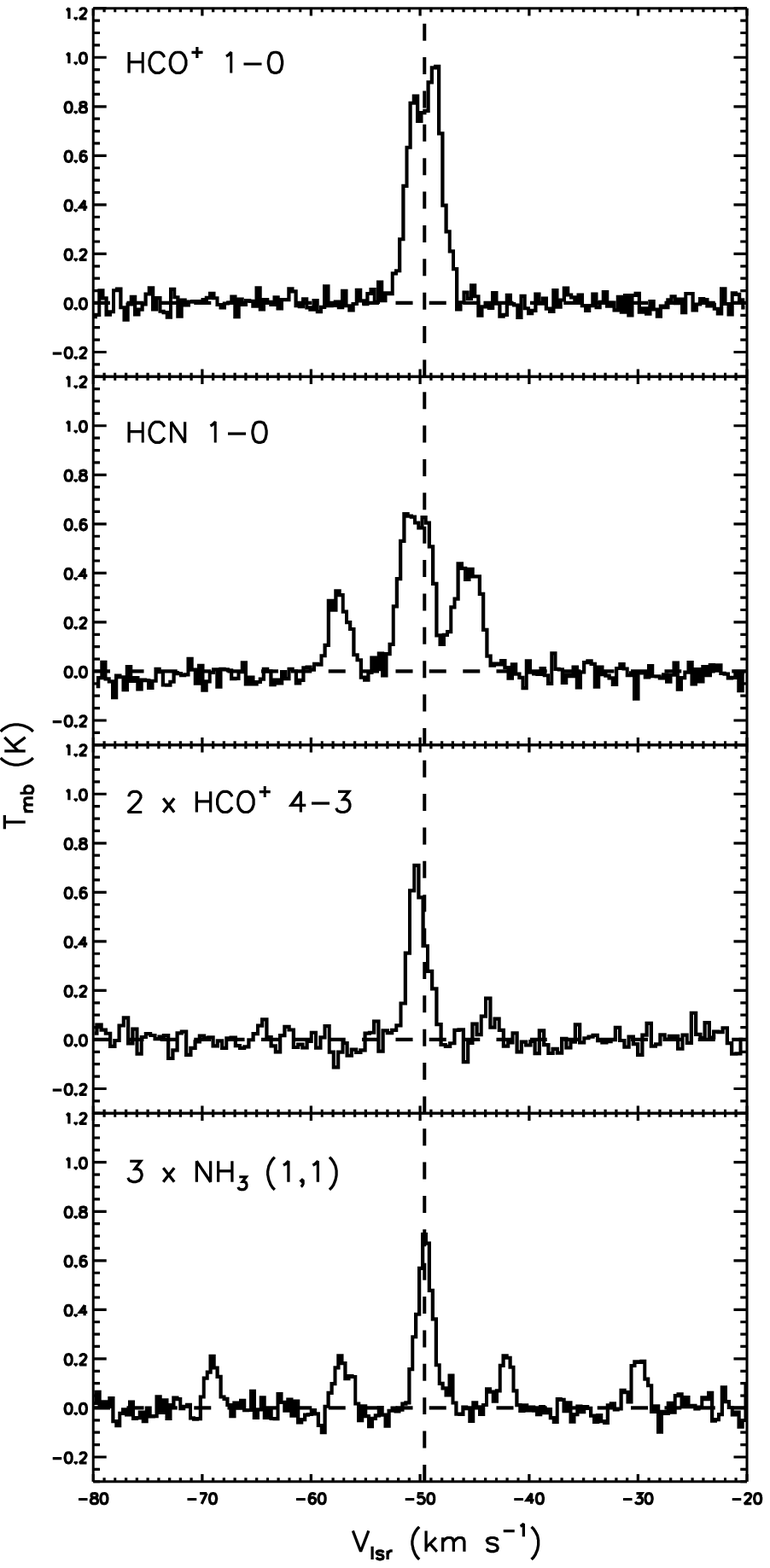}
\caption{Spectra of \hcop, \hcn, \hcopf, and \nht\ toward G108--N. }
\end{center}
\end{figure}

\begin{figure}[!p]
\epsscale{0.9}
\begin{center}
\plotone{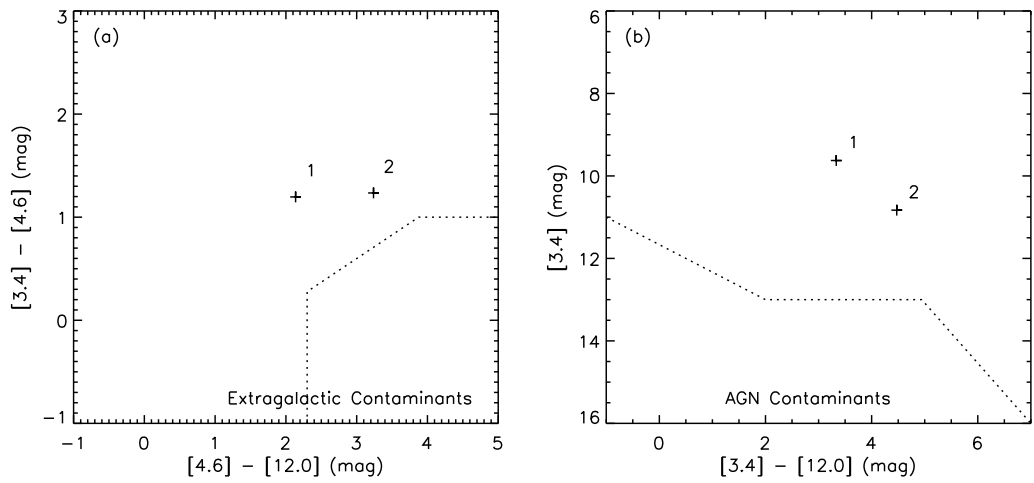}
\caption{(a) [3.4] - [4.6] versus [4.6] - [12.0] color-color diagram for IRS 1 and IRS 2.
The dashed lines describe the approximate criteria as likely star-forming galaxies \citep{koenig14}.
(b) [3.4] versus [3.4] - [12.0] color-magnitude diagram for IRS 1 and IRS 2.
The dashed lines indicates the region marked the approximate criteria as candidate AGNs \citep{koenig14}.
Source numbers are labeled.}
\end{center}
\end{figure}

\begin{figure}[!p]
\epsscale{0.6}
\begin{center}
\plotone{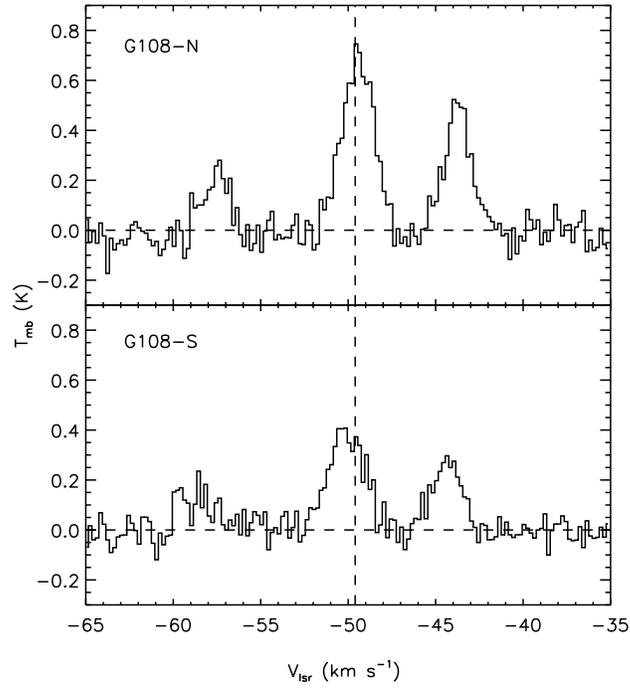}
\caption{The spectra of \nhp\ toward G108--N (\textit{upper}) and G108--S (\textit{lower}).}
\end{center}
\end{figure}
\clearpage

\begin{figure}[!p]
\epsscale{0.5}\begin{center}
\plotone{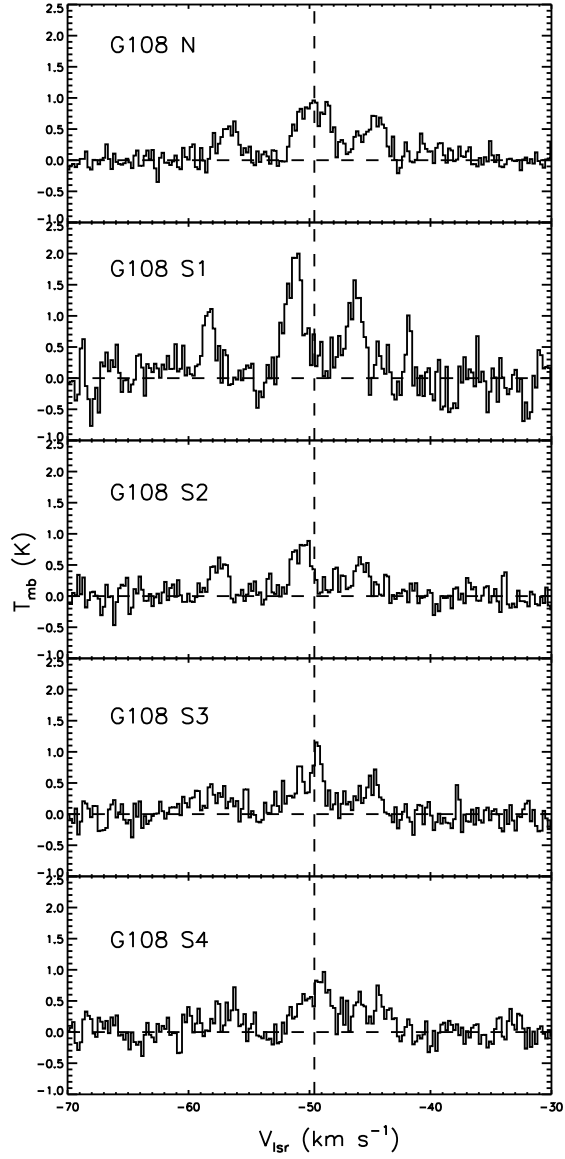}
\caption{Spectra of \hcn\ toward G108--N, G108--S1, G108--S2, G108--S3, and G108--S4 (\textit{top} to \textit{bottom}). }
\end{center}
\end{figure}

\begin{figure}[tbh!]
\centering
\includegraphics[angle=-90,scale=0.6]{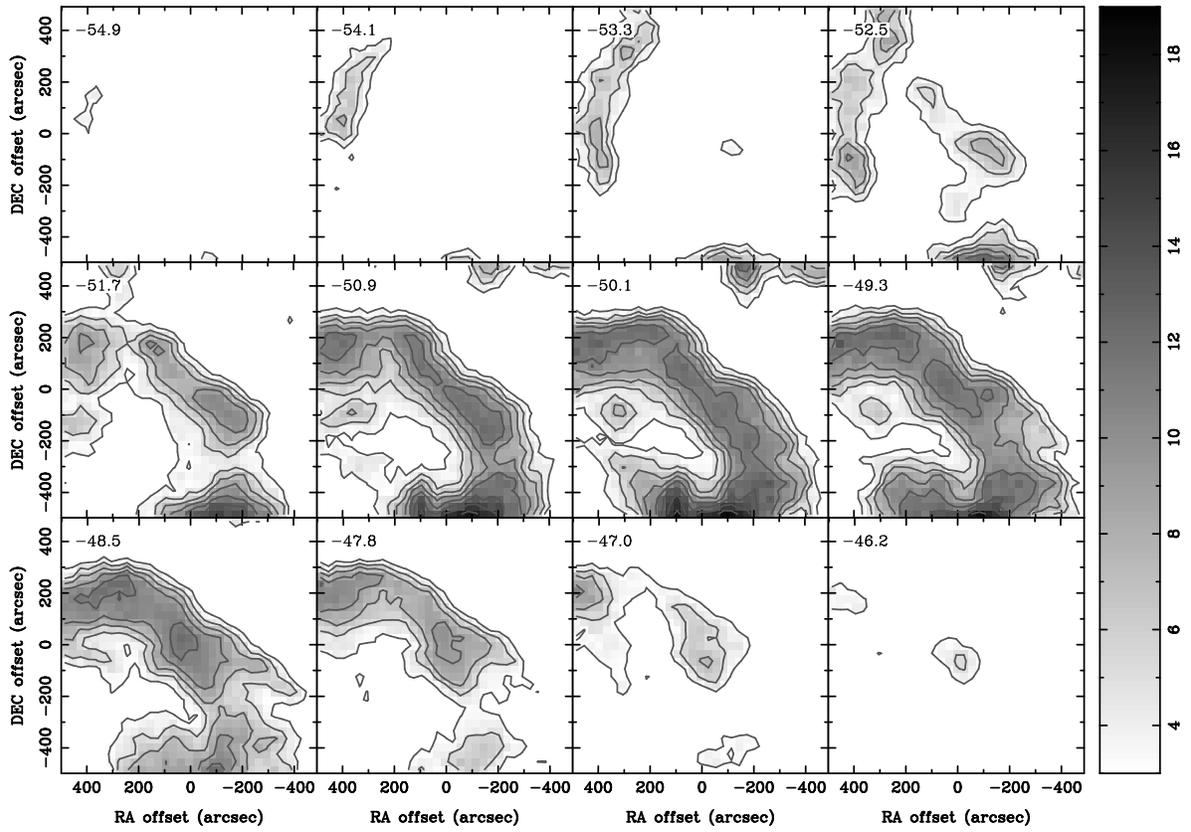}
\caption{Channel maps of \tco\ observed with PMO. Respective velocities are indicated in the upper left of each panel in km s$^{-1}$. 
The contours are from 3 to 19 in steps of 2 K km s$^{-1}$. 
The gray scale levels are from 3 to 19 in K km s$^{-1}$.}
\end{figure}

\clearpage


\begin{thebibliography}{99}

\bibitem[Andr{\'e} et al.(2014)]{andre14}
Andr{\'e}, P., Di Francesco, J., Ward-Thompson, D., et al.
2014, Protostars and Planets VI, 27

\bibitem[Battersby et al.(2011)]{battersby11} 
Battersby, C., Bally, J., Ginsburg, A., et al.
2011, \aap, 535, A128

\bibitem[Bourke et al.(2005)]{bourke05}
Bourke, T. L., Crapsi, A., Myers, P. C., et al.
2005, \apjl, 633, L129

\bibitem[Bintley et al. (2014)]{bint14}
Bintley, D., Holland, W. S., MacIntosh, M., J., et al. 
2014, \procspie, 9153, 3

\bibitem[Carey et al.(1998)]{carey98} 
Carey, S. J., Clark, F.~O., Egan, M.~P., et al.
1998, \apj, 508, 721

\bibitem[Caselli et al.(2002)]{caselli02}
Caselli, P., Benson, P.~J., Myers, P.~C., \& Tafalla, M.
2002, \apj, 572, 238

\bibitem[Chapin et al. (2013)]{chap13}
Chapin, E. L., Berry, D. S., Gibb, A. G., et al. 
2013, \mnras, 430, 2545

\bibitem[Dale et al.\ (2012)]{dale12}
Dale J. E., Ercolano B., Bonnell I. A.,
2012, \mnras, 427, 2852

\bibitem[Dale et al.\ (2015)]{dale15}
Dale, J. E., Haworth, T. J., Bressert, E.,
2015, \mnras, 450, 1199

\bibitem[Di Francesco et al.(2004)]{di04}
Di Francesco, J., Andr{\'e}, P., \& Myers, P. C.
2004, \apj, 617, 425

\bibitem[Dunham et al.(2006)]{dunham06}
Dunham, M. M., Evans, N. J., II, Bourke, T. L., et al.
2006, \apj, 651, 945

\bibitem[Elmegreen \& Lada(1977)]{el77}
Elmegreen, B. G., \& Lada, C. J.
1977, \apj, 214, 725

\bibitem[Evans(1999)]{evans99}
Evans, N. J., II
1999, \araa, 37, 311

\bibitem[Evans et al.(2009)]{evans09}
Evans, N. J., II, Dunham, M. M., J{\o}rgensen, J. K., et al.
2009, \apjs, 181, 321

\bibitem[Furuya et al.(2006)]{furuya06}
Furuya, R.~S., Kitamura, Y., \& Shinnaga, H.
2006, \apj, 653, 1369

\bibitem[Hatchell et al.(2013)]{hatchell13}
Hatchell, J., Wilson, T., Drabek, E., et al.
2013, \mnras , 429, L10

\bibitem[Hennebelle \& Chabrier(2008)]{hennebelle08}
Hennebelle, P., \& Chabrier, G.
2008, \apj, 684, 395

\bibitem[Heydari-Malayeri \& Testor(1981)]{heydari81} 
Heydari-Malayeri, M., \& Testor, G.
1981, \aap, 96, 219 

\bibitem[Hildebrand(1983)]{hildebrand83}
Hildebrand, R. H.
1983, \qjras , 24, 267

\bibitem[Ishihara et al.(2010)]{ishihara10}
Ishihara, D., Onaka, T., Kataza, H., et al.
2010, \aap, 514, A1

\bibitem[Kim et al.(2011)]{kim11}
Kim, K. T., Byun, D. Y., et al.
2011, Journal of Korean Astronomical Society, 44, 81

\bibitem[Koenig \& Leisawitz(2014)]{koenig14}
Koenig, X. P., \& Leisawitz, D. T.
2014, \apj, 791, 131

\bibitem[K{\"o}nyves et al.(2015)]{konyves15}
K{\"o}nyves, V., Andr{\'e}, P., Men'shchikov, A., et al.
2015, \aap, 584, A91

\bibitem[Krumholz \& McKee(2008)]{krumholz08} 
Krumholz, M. R., \& McKee, C. F.
2008, \nat, 451, 1082

\bibitem[Lada \& Lada(2003)]{lada03}
Lada, C. J., \& Lada, E. A.
2003, \araa , 41, 57

\bibitem[Lee et al.(2003)]{lee03}
Lee, J.-E., Evans, N. J., II, and Yancy, L. Shirley
2003, \apj, 583, 789

\bibitem[Liu et al.\ (2012a)]{liu12a}
Liu T., Wu Y., Zhang H., Qin S.-L.,
2012a, \apj, 751, 68

\bibitem[Liu et al.(2012)]{liu12}
Liu, T., Wu, Y., \& Zhang, H.
2012, \apjs, 202, 4

\bibitem[Liu et al.(2013)]{liu13}
Liu, T., Wu, Y., \& Zhang, H.
2013, \apjl, 775, L2

\bibitem[Liu et al.\ (2015)]{liu15}
Liu H.-L., Wu Y., Li J., Yuan J.-H., Liu T., Dong X.,
2015, \apj, 798, 30

\bibitem[Liu et al.(2016a)]{liu16a}
Liu, H.-L., Li, J.-Z., Wu, Y., et al.
2016a, \apj, 818, 95

\bibitem[Liu et al.(2016)]{liu16}
Liu, T., Zhang, Q., Kim, K.-T., et al.
2016, \apjs, 222, 7

\bibitem[Morgan et al.(2004)]{morgan04} 
Morgan, L. K., Thompson, M. A., Urquhart, J. S., et al. 
2004, \aap, 426, 535

\bibitem[Myers \& Ladd(1993)]{myers93}
Myers, P. C., \& Ladd, E. F.
1993, \apjl , 413, L47

\bibitem[Ossenkopf \& Henning(1994)]{ossenkopf94}
Ossenkopf, V., \& Henning, T.
1994, \aap, 291, 943

\bibitem[Pillai et al.(2006)]{pillai06}
Pillai, T., Wyrowski, F., Carey, S. J., et al.
2006, \aap, 450, 569

\bibitem[Planck Collaboration et al.(2011)]{planck11}
Planck Collaboration, Ade, P. A. R., Aghanim, N., et al.
2011, \aap, 536, A23

\bibitem[Planck Collaboration et al.(2015)]{planck15}
Planck Collaboration, Ade, P. A. R., Aghanim, N., et al.
2015, arXiv:1502.01599

\bibitem[Ram{\'{\i}}rez Alegr{\'{\i}}a et al.(2011)]{ramirez11}
Ram{\'{\i}}rez Alegr{\'{\i}}a, S., Herrero, A., Mar{\'{\i}}n-Franch, A., et al.
2011, \aap, 535, A8

\bibitem[Schnee et al.(2007)]{schnee07}
Schnee, S., Caselli, P., Goodman, A., et al.
2007, \apj, 671, 1839

\bibitem[Seta et al.(1998)]{seta98}
Seta, M., Hasegawa, T., Dame, T. M., et al.
1998, \apj, 505, 286

\bibitem[Simon et al.(2006)]{simon06}
Simon, R., Rathborne, J.~M., Shah, R.~Y., et al.
2006, \apj, 653, 1325

\bibitem[Su et al.(2014)]{su14}
Su, Y., Fang, M., Yang, J., et al.
2014, \apj, 788, 122

\bibitem[Tafalla et al.(2002)]{tafalla02}
Tafalla, M., Myers, P. C., Caselli, P., et al.
2002, \apj, 569, 815

\bibitem[Tatematsu et al.(1987)]{tatematsu87}
Tatematsu, K., Fukui, Y., Nakano, M., et al.
1987, \aap, 184, 279

\bibitem[Tatematsu et al.(1990)]{tatematsu90}
Tatematsu, K., Fukui, Y., Iwata, T., et al.
1990, \apj, 351, 157

\bibitem[Tatematsu et al.(2017)]{tatematsu17} 
Tatematsu, K., Liu, T., Ohashi, S., et al.
2017, \apjs, 228, 12

\bibitem[Thompson et al.\ (2012)]{thom12}
Thompson, M. A., Urquhart, J. S., Moore, T. J. T., et al.
2012, \mnras, 421, 408

\bibitem[Ungerechts et al.(2000)]{ungerechts00}
Ungerechts, H., Umbanhowar, P., \& Thaddeus, P.
2000, \apj, 537, 221

\bibitem[Urquhart et al.(2014)]{urquhart14} 
Urquhart, J.~S., Moore, T.~J.~T., Csengeri, T., et al.
2014, \mnras, 443, 1555 

\bibitem[van der Tak et al.(2007)]{vandertak07}
van der Tak, F. F. S., Black, J. H., Sch{\"o}ier, F. L., et al.
2007, \aap, 468, 627

\bibitem[Whitworth et al.\ (1994)]{whit94}
Whitworth, A. P., Bhattal, A. S., Chapman, S. J., et al.
1994, \mnras, 268, 291


\bibitem[Wright et al.(2010)]{wright10}
Wright, E. L., Eisenhardt, P. R. M., Mainzer, A. K., et al.
2010, \aj, 140, 1868


\bibitem[Wu et al.(2012)]{wu12}
Wu, Y., Liu, T., Meng, F., et al.\
2012, \apj, 756, 76

\bibitem[Young et al.(2004)]{young04}
Young, C. H., J{\o}rgensen, J. K., Shirley, Y. L., et al.
2004, \apjs, 154, 396

\bibitem[Yung et al.(2014)]{yung14}
Yung, B. H. K., Nakashima, J. i., \& Henkel, C.
2014, \apj, 794, 81

\bibitem[Zhang et al.(2016)]{zhang16}
Zhang, T., Wu, Y., Liu, T., el al.
2016, \apjs, 224, 43

\end{thebibliography}
\end{document}